\documentclass[11pt,prd,preprintnumbers,amsmath,amssymb,nofootinbib]{article}
\pdfoutput=1

\usepackage[utf8]{inputenc}
\usepackage{amsmath,amssymb,amscd}
\usepackage{listings}
\usepackage{caption}
\usepackage{dsfont}
\usepackage{slashed}
\usepackage{color}
\usepackage[caption=false]{subfig}
\usepackage[pdftex]{graphicx}
\usepackage{epstopdf}
\usepackage{listings}
\usepackage{epsfig}
\usepackage{listings}
\usepackage[font=small,labelfont=bf]{caption}
\usepackage{cite}
\usepackage{hyperref}
\usepackage{paralist}
\usepackage{placeins}
\usepackage{graphicx}
\usepackage{enumitem}
\usepackage{bm}
\usepackage{xcolor,colortbl}
\usepackage{multirow}

\usepackage{subfig}
\usepackage{color,graphicx,slashed,hyperref}
\usepackage[utf8]{inputenc}
\hypersetup{
   bookmarks=true,         
   unicode=true,          
   pdftoolbar=true,        
   pdfmenubar=true,        
   pdffitwindow=false,     
   pdfstartview={FitH},    
   pdftitle={Draft},    
   pdfauthor={Author},     
   pdfsubject={Subject},   
   pdfcreator={Creator},   
   pdfproducer={Producer}, 
   pdfkeywords={keyword1} {key2} {key3}, 
   pdfnewwindow=true,      
   colorlinks=true,       
   linkcolor=blue,          
   citecolor=blue,        
   filecolor=black,      
   urlcolor=blue           
}

\selectfont

\newcommand{\beq} {\begin{equation}}
\newcommand{\eeq} {\end{equation}}
\newcommand{\bea} {\begin{eqnarray}}
\newcommand{\eea} {\end{eqnarray}}
\newcommand{\ba} {\begin{eqnarray*}}
\newcommand{\ea} {\end{eqnarray*}}

\definecolor{Celadon}{rgb}{0.67, 0.94, 0.82}
\definecolor{Pink}{rgb}{0.9, 0.0, 0.0}
\definecolor{darkred}{rgb}{0.5, 0.2, 0.13}
\definecolor{ForestGreen}{RGB}{34,139,34}

\renewenvironment{align}{
    \begin{equation}
    \begin{aligned}
}{
    \end{aligned}
    \end{equation}
    \ignorespacesafterend
}

\begin{document}

\thispagestyle{empty} 
\begin{flushright}
TTP21-009\\ 
P3H-21-025
\end{flushright}
\vspace*{1.cm}\par
\begin{center}	
{\par\centering \textbf{  
\LARGE \bf Muon $g-2$ and $B$-anomalies from Dark Matter}} \\
\vskip 1.cm\par
{\scalebox{.88}{\par\centering \large  
\sc Giorgio Arcadi$^{\,1,2}$, 
~Lorenzo Calibbi$^{\,3}$,  
~Marco Fedele$^{\,4}$, 
and Federico Mescia$^{\,5}$}} \\
{\par\centering \vskip 0.6 cm\par}
{\small
$^1$ \textit{Dipartimento di Matematica e Fisica, Universit\`a di Roma 3, Via della Vasca Navale 84, 00146, Roma, Italy}\\
$^2$ \textit{INFN Sezione Roma 3}\\
$^3$  \textit{School of Physics, Nankai University, Tianjin 300071, China} \\
$^4$  \textit{Institut f\"ur Theoretische Teilchenphysik, Karlsruhe Institute of Technology, D-76131 Karlsruhe, Germany} \\
$^5$  \textit{Dept.~de F\'{\i}sica Qu\`antica i Astrof\'{\i}sica, Institut de Ci\`encies del Cosmos (ICCUB), Universitat de Barcelona,\\Mart\'i i Franqu\`es 1, E-08028 Barcelona, Spain} } \\
{\vskip 1.cm\par}
\begin{abstract}
\noindent 
In the light of the recent result of the Muon g-2 experiment and the update on the test of lepton flavour universality $R_K$ published by the LHCb collaboration,
we systematically build and discuss a set of models with minimal field content that can simultaneously give: (i) a thermal Dark Matter candidate; (ii) large loop contributions to $b\to s\ell\ell$ processes able to address $R_K$ and the other $B$ anomalies; (iii) a natural solution to the muon $g-2$ discrepancy through chirally-enhanced contributions.
\end{abstract}
{\vskip 9.cm\par}
\emph{E-mail:} 
\href{mailto:giorgio.arcadi@uniroma3.it}{giorgio.arcadi@uniroma3.it}, \href{mailto:calibbi@nankai.edu.cn}{calibbi@nankai.edu.cn}, \href{mailto:marco.fedele@kit.edu}{marco.fedele@kit.edu}, \href{mailto:mescia@ub.edu}{mescia@ub.edu} 
\end{center}

\newpage

\setcounter{footnote}{0}

\newenvironment{Appendix}
{
	\setcounter{section}{1}
	\setcounter{equation}{0}
	\renewcommand{\thesubsection}{\Alph{subsection}}
	\renewcommand{\theequation}{A.\arabic{equation}}
}


\section{Introduction}
The first results of the FNAL Muon g-2 experiment~\cite{Abi:2021gix} have confirmed the long-standing discrepancy with the Standard Model (SM) prediction of the anomalous magnetic moment of the muon $a_\mu \equiv (g-2)_\mu/2$:
\beq
\Delta a_\mu \equiv a^\text{exp}_\mu - a^\text{SM}_\mu
= 251(59) \times 10^{-11} \,.
\eeq
The above deviation between measurement and theoretical prediction amounts to about $4.2\sigma$, and takes into account the combination with the previous measurement of the BNL experiment~\cite{Bennett:2006fi}, drastically reducing the probability of a statistical fluctuation or overlooked systematical effects.\footnote{The deviation obtained taking into account only the FNAL data amounts to about $3.3\sigma$.} It is also unlikely that such a discrepancy can be fully explained by underestimated hadronic uncertainties~\cite{Aoyama:2020ynm}. Moreover, even if hadronic vacuum polarization effects are assumed to be large enough to account for the anomaly~\cite{Borsanyi:2020mff}, this would cause a deterioration of the electroweak~(EW) fit such that tensions of comparable significance in other EW observables would arise~\cite{Passera:2008jk,Crivellin:2020zul,Keshavarzi:2020bfy,Malaescu:2020zuc}.
Hence this new result strongly supports the case for new physics~(NP) requiring, in particular, the presence of new particles with non-trivial interactions with SM muons at scales $\lesssim 100$~TeV~\cite{Capdevilla:2020qel,Buttazzo:2020eyl,Capdevilla:2021rwo}, where the upper bound can be reached only in presence of fields strongly coupled with muons, and at the price of fine-tuned cancellations between SM and NP contributions to the muon mass.

Interestingly, also the persistent anomalies in semileptonic $B$ meson decays of the kind $b\to s\ell\ell$ seem to point to a NP sector with preferred couplings to muons. In particular, the theoretically
clean lepton flavour universality (LFU) ratio
$R_K = \text{BR}(B\to K\mu^+\mu^-)/\text{BR}(B\to K e^+ e^-)$, for which an updated measurement including the full Run I + Run II dataset has been recently released by the LHCb collaboration,
deviates from the SM prediction by more than $3\sigma$~\cite{Aaij:2021vac}.\footnote{Recent interpretations of this measurement can be found in Refs.~\cite{Angelescu:2021lln,Hiller:2021pul,Greljo:2021xmg,Cornella:2021sby,Kriewald:2021hfc,Carvunis:2021jga}.}
Once the LFU ratio $R_{K^*} = \text{BR}(B\to K^*\mu^+\mu^-)/\text{BR}(B\to K^* e^+ e^-)$~\cite{Aaij:2017vbb,Abdesselam:2019wac} and the branching ratios and angular analysis of other decays mediated by $b\to s\ell\ell$ transitions~\cite{Aaij:2015dea,Aaij:2015esa,Khachatryan:2015isa,Aaij:2015oid,Aaij:2016flj,Wehle:2016yoi,Sirunyan:2017dhj,Aaboud:2018krd,Aaij:2020nrf,Aaij:2020ruw} are considered as well, global fits to data prefer the presence
of NP contributions at the level of $\gtrsim 5\sigma$~\cite{DAmico:2017mtc,Ciuchini:2019usw,Alguero:2019ptt,Alok:2019ufo,Datta:2019zca,Aebischer:2019mlg,Kowalska:2019ley,Ciuchini:2020gvn,Hurth:2020ehu,Altmannshofer:2021qrr,Geng:2021nhg} compared to the SM prediction only.
These anomalies could also be explained by new particles interacting with muons at scales $\lesssim 100$~TeV~\cite{DiLuzio:2017chi}. It is therefore extremely tempting to discuss NP models that can provide a common explanation of the muon $g-2$ discrepancy and the $B$-physics anomalies.

Motivated by the overwhelming evidence for Dark Matter (DM) in the universe~\cite{Bertone:2016nfn}, which is perhaps the strongest call for physics beyond the SM (BSM), the aim of this paper is to show how the two anomalies can arise by loops involving the very same fields of the DM sector, including a thermal DM candidate.
The idea is to build a set of models with minimal field content
that can simultaneously account for the anomalies due to interactions between the DM fields, other NP particles, and SM fermions (muons, bottom and strange quarks). Dark Matter stability requires 
that the couplings of interactions involving two SM fields and the DM field are very suppressed. For definiteness, we assume that 
such interactions are forbidden by a global (possibly accidental) symmetry (discrete or continuous), whose other effect is to 
prohibit mixing between SM and NP fields.
Under these assumptions NP contributions to both $a_\mu$
and $b\to s\mu^+\mu^-$ can only occur through loop diagrams, as in the framework discussed in Refs.~\cite{Gripaios:2015gra,Arnan:2016cpy,Arnan:2019uhr},
where only NP fields (DM in particular) run in the loop.
Models of this kind for DM, the $g-2$ and/or the $B$ anomalies have been discussed in Refs.~\cite{Belanger:2015nma,Arcadi:2017kky,Kawamura:2017ecz,Kowalska:2017iqv,Cline:2017qqu,Calibbi:2018rzv,Barman:2018jhz,Grinstein:2018fgb,Cerdeno:2019vpd,Okada:2019sbb,Mohan:2019zrk,Guadagnoli:2020tlx,Carvunis:2020exc,Biswas:2019twf,Calibbi:2019bay,Darme:2020hpo,Kawamura:2020qxo,Huang:2020ris,Okawa:2020jea,Kowalska:2020zve,DEramo:2020sqv,Arcadi:2021glq,Becker:2021sfd,Yin:2021yqy}. In particular, in Ref.~\cite{Arcadi:2021glq}, we systematically built and studied the minimal models that, by introducing three NP fields only, can simultaneously explain DM and the $B$-anomalies. Based on the results of our previous work, here 
we show the minimal ingredients required by models where
also the muon $g-2$ anomaly is naturally accounted for.

\begin{figure}[!t]
\centering
\includegraphics[width=0.90\textwidth]{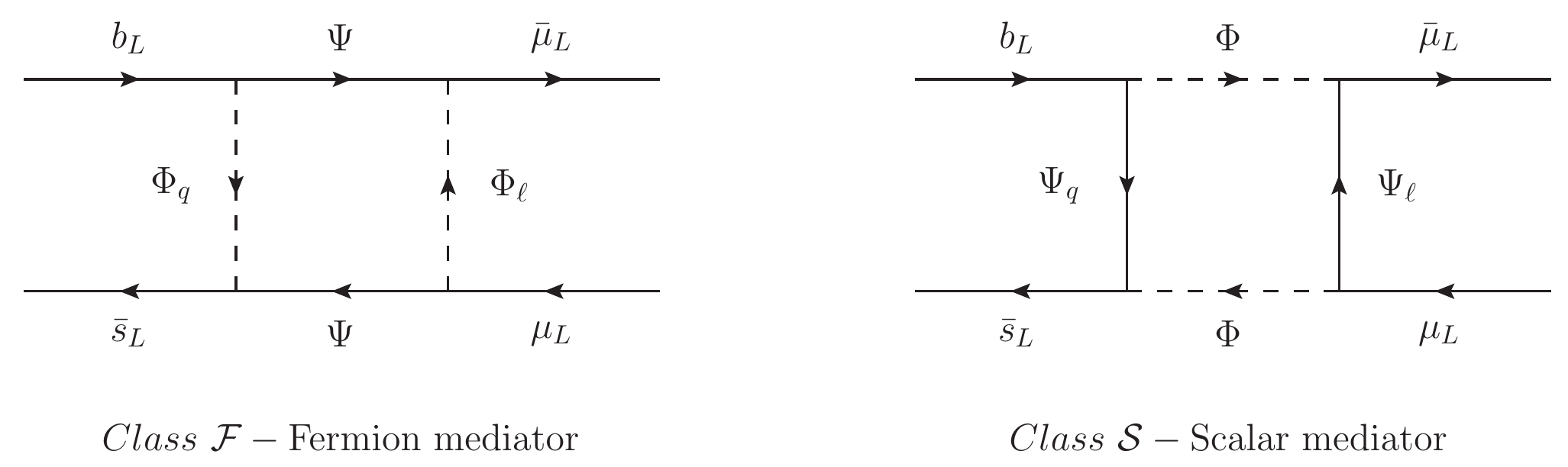}
\caption{Basic diagrams providing a contribution to $b\to s\mu\mu$
involving only left-handed SM fields, i.e.~of the kind $\delta C^{9}_\mu=-\delta C^{10}_\mu$.
Based on this, we classify the models according to the spin
of the \emph{flavour mediator}, the field that couple to both quarks and leptons.
\label{fig:boxes}}
\end{figure}


\section{Setup}
\label{sec:setup}
Recent global analyses of the $b\to s\ell\ell$ data~\cite{DAmico:2017mtc,Ciuchini:2019usw,Alguero:2019ptt,Alok:2019ufo,Datta:2019zca,Aebischer:2019mlg,Kowalska:2019ley,Ciuchini:2020gvn,Hurth:2020ehu},
including the new measurement of $R_K$~\cite{Altmannshofer:2021qrr,Geng:2021nhg}, show
that  a satisfactory fit of the observed $B$-anomalies favours solutions featuring effects in $\delta C^{9}_\mu$ and $\delta C^{10}_\mu$, where these quantities are defined as the NP contributions to the following operators:
\begin{align}
\mathcal{H}_{\rm eff} ~\supset~ -\frac{4 G_F}{\sqrt{2}} \frac{e^2}{16\pi^2}V_{tb} V_{ts}^* \left[C^9_\mu \,(\overline{s}\gamma_\mu P_L b) (\overline{\mu} \gamma^\mu\mu)+ C^{10}_\mu\, (\overline{s}\gamma_\mu P_L b) (\overline{\mu} \gamma^\mu\gamma_5 \mu)	  +{\rm h.c.} \right].
\label{eq:Heff}
\end{align}
In particular, the simplest ways to improve the fit to the data is to introduce an exotic contribution to $\delta C^{9}_\mu$ alone or one of the kind $\delta C^{9}_\mu=-\delta C^{10}_\mu$.
Hence substantial (or exclusive) interactions involving left-handed (LH) muons are favoured. Furthermore, global fits require that non-standard contributions from hadronic right-handed (RH) currents (if present at all) be subdominant. 
In other words, a minimal ingredient of our models will be given
by the 1-loop contributions shown in Figure~\ref{fig:boxes}, that is, the three fields appearing in either diagram need to be present and one of them will constitute the DM candidate, as discussed in our previous work~\cite{Arcadi:2021glq}.

Our previous analysis showed that the most satisfactory solutions of the $B$ anomalies (that is, the only viable ones in wide regions of the parameter space without relying on tuning) that provide in addition a natural thermal DM candidate require: (i) DM to be an $SU(2)_L\times U(1)_Y$ singlet; (ii) the DM field to couple to muons (since the large couplings to muons required by the fit to the $B$-anomalies induce efficient DM annihilation evading the problem of DM overproduction); (iii) DM to be a Majorana fermion, a real scalar, or one of the two components of a complex scalar with a mass splitting $>\mathcal{O}(100)$~keV (such that the most dangerous contributions to DM direct detection are suppressed).
The above considerations restrict the set of viable possibilities to cases where the fields $\Psi/\Phi$ or $\Phi_\ell/\Psi_\ell$ in Figure~\ref{fig:boxes} are (or mix with) a DM singlet. 

\begin{figure}[!t]
\centering
\includegraphics[width=0.45\textwidth]{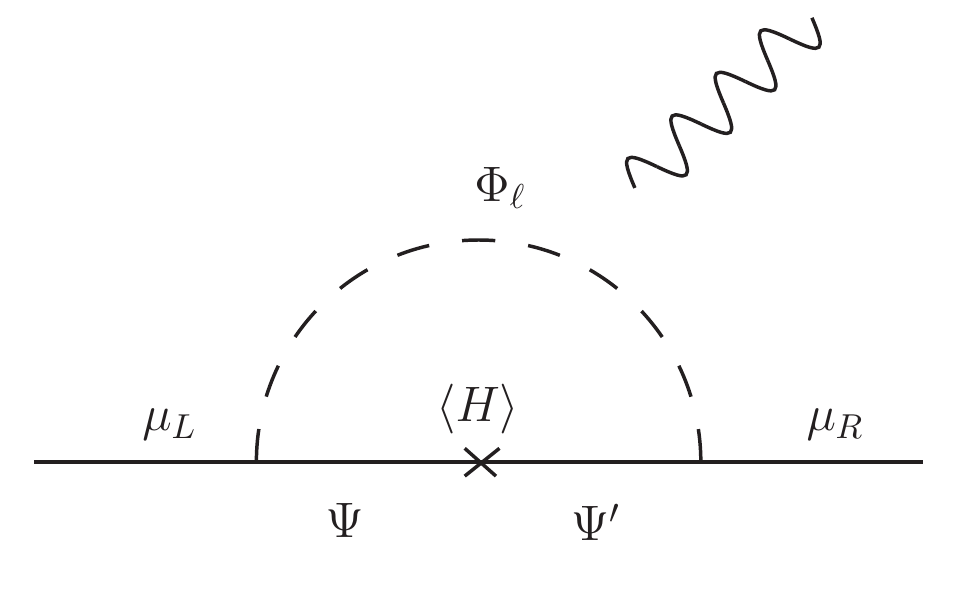}
\hfill
\includegraphics[width=0.45\textwidth]{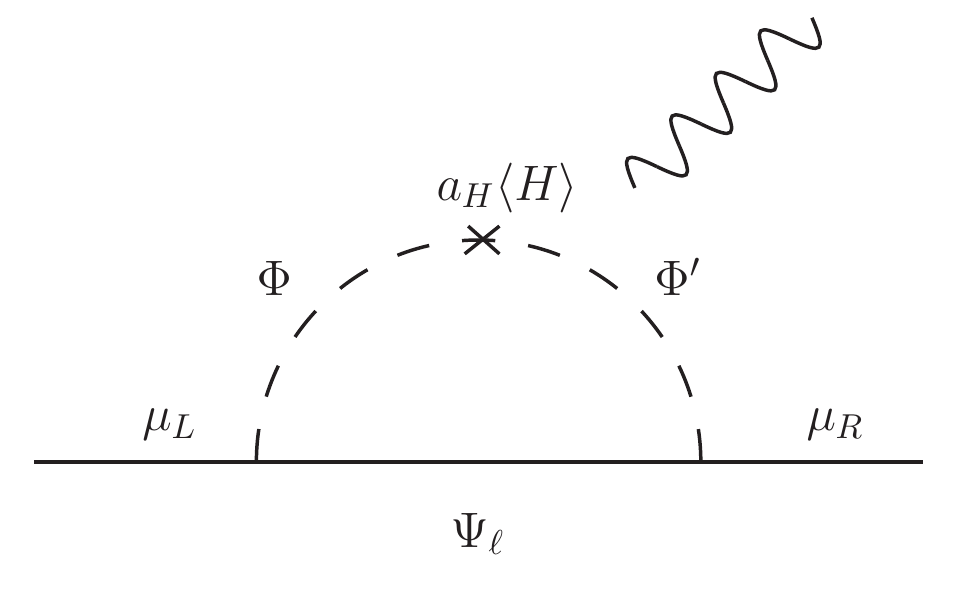}
\caption{Diagrams giving chirally-enhanced contributions to $(g-2)_\mu$.
\label{fig:g-2}}
\end{figure}

\begin{figure}[!t]
\centering
\includegraphics[width=0.90\textwidth]{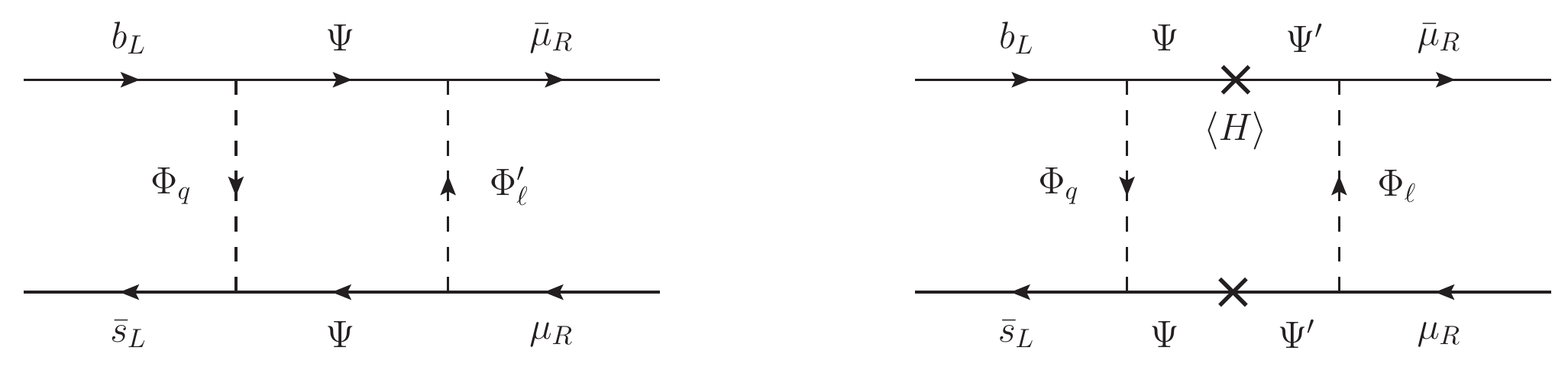}
\caption{Additional diagrams contributing to $b\to s\mu\mu$ involving RH muons.
\label{fig:new-boxes}}
\end{figure}

The subset of NP fields coupling to muons in Figure~\ref{fig:boxes}
also contributes to the dipole operator relevant for the
muon $g-2$:
\beq
    {\cal L} ~\supset~ \frac{e\, v}{8\pi^2}\, C_{\mu\mu}\left(\bar\mu_L \sigma_{\mu \nu} \mu_R\right)\, F^{\mu\nu} + {\rm h.c.} 
\quad \Rightarrow \quad \Delta a_\mu = \frac{{{m_{\mu}}v}}{2 \pi^2}\, \mbox{\rm Re} (C_{\mu \mu}),
\eeq 
where $v$ is the SM Higgs vacuum expectation value (vev) $\simeq 246$~GeV. The normalisation $\propto v$ of the above operator highlights that, following from gauge invariance, 
a flip of the chirality of the muon, hence a Higgs vev insertion, is necessary to induce such effect. On the other hand, the fields
$\Psi-\Phi_\ell$ or $\Phi-\Psi_\ell$ do not couple to RH muons, hence such a chirality flip can only occur through a muon mass insertion in the external leg, leading to a suppression of the effect by the small muon Yukawa coupling, $C_{\mu\mu}\propto y_\mu$. Minimal models where DM couples only to LH muons therefore can not provide a sizeable contribution to $a_\mu$, besides very tuned 
regions of the parameter space~\cite{Calibbi:2018rzv}.
Thus, a natural explanation of the muon $g-2$ anomaly requires a chiral enhancement, i.e.~a chirality flip occurring inside the loop through a coupling to the SM Higgs field $\gg y_\mu$, see e.g.~\cite{Kowalska:2017iqv,Calibbi:2018rzv,Crivellin:2018qmi,Arnan:2019uhr}. The minimal way to achieve this is to add a 4th field to our minimal models: either $\Psi^\prime/\Phi^\prime$ mixing with $\Psi/\Phi$ through a Higgs vev, or $\Phi^\prime_\ell/\Psi^\prime_\ell$ mixing with $\Phi_\ell/\Psi_\ell$. Illustrative diagrams providing an enhanced contribution to the muon $g-2$ are shown in Figure~\ref{fig:g-2}.
Notice that these mixing fields also induce additional contributions to $b\to s\mu\mu$ involving RH muons (thus deviating from the $\delta C^{9}_\mu=-\delta C^{10}_\mu$ pattern), as shown in Figure~\ref{fig:new-boxes}.

The only combinations of the quantum numbers of the new fields that fulfils the above conditions are displayed in Table~\ref{tab:reps}. A unique choice of the transformation properties under $SU(3)_c$ and only three under $SU(2)_L$ are possible. For each of these three choices a minimal model would comprise four fields:  
\bea
\text{Class }\mathcal{F}:&\quad&\text{either}\quad\{\Phi_q,\,\Phi_\ell,\,\Phi^\prime_\ell,\,\Psi\}\quad\text{or}\quad\{\Phi_q,\,\Phi_\ell,\,\Psi,\,\Psi^\prime\}
\\
\text{Class }\mathcal{S}:&\quad&\text{either}\quad\{\Psi_q,\,\Psi_\ell,\,\Psi^\prime_\ell,\,\Phi\}\quad\text{or}\quad\{\Psi_q,\,\Psi_\ell,\,\Phi,\,\Phi^\prime\}
\eea

Considering the two possible choices of the mixing field, as well as the possible hypercharge assignments delivering at least an absolute singlet coupling to leptons, we end up with only 9 options (times the two spin alternatives). These are listed in Table~\ref{tab:models}.
The models highlighted in cyan feature pure singlet DM (scalar or fermion); for the models highlighted in red, DM is in general a mixture of a singlet and an $SU(2)_L$ doublet (again scalar or fermion).
\begin{table}[t]
\centering
\renewcommand{\arraystretch}{1.1}
\begin{tabular}{ | c | c  c  c |c|c|  }
\hline
${SU\left( 3 \right)_c}$ & ${{\Phi_q}/{\Psi_q}}$ & ${{\Phi_\ell}/{\Psi_\ell}}$ & ~~${\Psi /\Phi }$~~~ &
~${{\Phi^\prime_\ell}/{\Psi^\prime_\ell}}$~ & 
${\Psi^\prime /\Phi^\prime }$ \\
\hline
\rowcolor{green!50}
   A &\bf 3&\bf 1&\bf 1  &\bf 1  &\bf 1  \\
   B &\bf 1& ${\bf \bar 3}$&\bf 3 & ${\bf \bar 3}$&\bf 3 \\
\hline
\hline
${SU\left( 2 \right)_L}$ & ${{\Phi_q}/{\Psi_q}}$ & ${{\Phi_\ell}/{\Psi_\ell}}$ & $ {\Psi /\Phi}$ &${{\Phi^\prime_\ell}/{\Psi^\prime_\ell}}$ & ${\Psi^\prime /\Phi^\prime }$\\
\hline
\rowcolor{green!50}
   I & \bf 2 &\bf 2  &\bf 1 &\bf 1 &${\bf 2 }$\\
\rowcolor{green!50}
  II & \bf 1 &\bf 1 &\bf 2 &${\bf 2 }$ &\bf 1\\
  III & \bf 3 &\bf 3 &\bf 2 &${\bf 2 }$ &\bf 3\\
  IV & \bf 2 &\bf 2 &\bf 3 &\bf 3 &${\bf 2 }$\\
\rowcolor{green!50}
    V & \bf 3 &\bf 1 &\bf 2 &${\bf 2 }$ &\bf 1\\
   VI & \bf 1 &\bf 3 &\bf 2 &${\bf 2 }$ &\bf 3\\
\hline
\hline
$U(1)_Y$ & ${{\Phi_q}/{\Psi_q}}$ & ${{\Phi_\ell}/{\Psi_\ell}}$ & $ {\Psi /\Phi}$ &${{\Phi^\prime_\ell}/{\Psi^\prime_\ell}}$ & ${\Psi^\prime /\Phi^\prime }$\\
\hline
    & $1/6-X$ & $-1/2-X$ & $X$  &  $-1-X$ & $-1/2 + X$\\
\hline  
\end{tabular}
\caption{Possible gauge quantum numbers of the new fields. 
Our convention for the hypercharge ($Q = Y +T_3$) is such that the SM fields have $Y(Q)=1/6$, $Y(U)=2/3$, $Y(D)=-1/3$, $Y(L)= -1/2$, $Y(E)= -1$, $Y(H)= 1/2$.
We highlight in green the combinations that provide a viable simultaneous fit to DM and $B$-anomalies. Minimal models includes the first three fields plus one chosen from the last two columns. 
\label{tab:reps}}
\end{table}


\section{Minimal models}
\label{sec:models}
In the previous section, we showed how our set of phenomenological requirements lead to a limited number of minimal models featuring four NP fields, which are displayed in Table~\ref{tab:models}. Here we provide more details about the interactions and the field mixing occurring in the different cases. We classify the models according to the nature of the ``flavour mediator'' field appearing in the diagrams of Figure~\ref{fig:boxes}.
\paragraph{Class $\mathcal{F}$.} These models feature a vector-like fermion $\Psi$ as flavour mediator and two extra scalars $\Phi_q$ and $\Phi_\ell$ coupling to the SM left-handed fermions. For models augmented with 
an additional scalar $\Phi_\ell^\prime$, the interactions are described by the following Lagrangian:
\begin{align}
{\cal L}^{\Phi_\ell \Phi_\ell^\prime}_\mathcal{F}~ \supset ~ \Gamma^Q_i\, \bar Q_i\, P_R \Psi \, \Phi_q  + \Gamma^L_i \, \bar L_i\, P_R \Psi \, \Phi_\ell  + \Gamma^E_i \, \bar E_i\, P_L \Psi \, \Phi^\prime_\ell  +
 a_H \Phi^\dag_\ell \Phi_\ell^\prime\,H
 +{\rm{h}}{\rm{.c.}}\,,
 \end{align}
where $a_H$ is a parameter with a dimension of a mass. In case of $\Phi_\ell^{\prime}$ being a doublet, one may need to substitute $\Phi_\ell^{\prime}\to \widetilde \Phi_\ell^{\prime} = i\sigma_2 \Phi_\ell^{\prime}$.
Notice that, depending on the hypercharge,  either the charged or the neutral components in $\Phi_\ell$ and $\Phi_\ell^{\prime}$ mix upon EW-symmetry breaking (EWSB). The resulting mass eigenstates and the corresponding mass eigenvalues will be given by diagonalising
a matrix of the following schematic form:
\begin{equation}
\mathcal{M}^2_\Phi = \left(
\begin{array}{cc}
   M_{\Phi_\ell}^2   & {a_H v}/{\sqrt{2}} \\
      {a^*_H v}/{\sqrt{2}} & M_{\Phi_\ell^\prime}^2 
\end{array}
\right)\,.
\end{equation}
For models where instead the fourth field is the additional fermion $\Psi^\prime$ mixing with the flavour mediator $\Psi$,  the Lagrangian schematically reads:
\begin{align}
{\cal L}^{\Psi \Psi^\prime}_\mathcal{F}~ \supset ~ \Gamma^Q_i\, \bar Q_i\, P_R \Psi \, \Phi_q  + \Gamma^L_i \,  \bar L_i\, P_R  \Psi \, \Phi_\ell  + \Gamma^E_i \, \bar  E_i\, P_L \Psi^\prime \, \Phi_\ell    + \lambda_{HL}  \bar\Psi\,P_{L}  \Psi^\prime\,H
+  {\lambda}_{HR}   \bar \Psi P_R  \Psi^\prime   {H} + {\rm{h.c.}}\,,
 \end{align}
For illustration here we show the case labelled $\mathcal{F}_\text{IIc}$ (or equivalently $\mathcal{F}_\text{Vc}$) in Table~\ref{tab:models}, where $\Psi$ is a doublet and $\Psi^\prime$ a Majorana or Dirac singlet (we recall all the extra fermions, unless they are Majorana, come in vectorlike pairs).
We have also omitted couplings to RH quarks, possibly allowed by gauge invariance,
of the kind $\Gamma^D_i \bar D_i\,P_R \Psi^\prime \, \Phi_q$ and $\Gamma^U_i  \bar U_i\,P_R \Psi^\prime \, \Phi_q$
(that we are assuming to be suppressed in the following).\footnote{Notice that models in the categories $\mathcal{F_\text{V}}/\mathcal{S_\text{V}}$ are identical to those in $\mathcal{F_\text{II}}/\mathcal{S_\text{II}}$ besides the field $\Phi_q/\Psi_q$ being an $SU(2)_L$ triplet instead of a singlet. This forbids coupling to RH quarks but we expect that otherwise it is not modifying DM and flavour phenomenology to large extent. Thus we can omit a detailed analysis of this class models and focus on those belonging to $\mathcal{F_\text{I}}/\mathcal{S_\text{I}}$ and $\mathcal{F_\text{II}}/\mathcal{S_\text{II}}$.}
For this kind of models the singlet-doublet mass matrix has the schematic forms:
 \begin{equation}
 \mathcal{M}^M_\Psi = \left(
 \begin{array}{ccc}
   M_{\Psi^\prime}   &  \lambda_{HL}v/{\sqrt{2}} & \lambda_{HR}^*v/{\sqrt{2}} \\
     \lambda_{HL}v/{\sqrt{2}} & 0 & M_{\Psi}\\
     \lambda_{HR}^*v/{\sqrt{2}} & M_{\Psi} & 0\\
 \end{array}
 \right)\,,\quad\quad
 \mathcal{M}^D_\Psi = \left(
 \begin{array}{cc}
   M_{\Psi^\prime}   &  \lambda_{HR}^*v/{\sqrt{2}}  \\
     \lambda_{HL}v/{\sqrt{2}} & M_{\Psi}\\
 \end{array}
 \right)\,,
\end{equation}
for, respectively, a Majorana and a Dirac singlet field ($\Psi^\prime$ in this illustrative examples).

\begin{table}[t!]
\centering
\renewcommand{\arraystretch}{1.4}
\begin{tabular}{ | c | c  c  c  c c|  }
\hline
Label & $\Phi_q/\Psi_q$ & $\Phi_\ell/\Psi_\ell$ & $\Psi/\Phi$ & $\Phi^\prime_\ell/\Psi^\prime_\ell$ & $\Psi^\prime/\Phi^\prime$ \\
\hline\hline
\rowcolor{cyan!40}
$\mathcal{F}_\text{Ia}/\mathcal{S}_\text{Ia}$
& $({\bf 3},{\bf 2},1/6)$ & $({\bf 1},{\bf 2},-1/2)$ & $({\bf 1},{\bf 1},0)$ & $({\bf 1},{\bf 1},-1)$ &  -- \\
\rowcolor{red!35}
$\mathcal{F}_\text{Ib}/\mathcal{S}_\text{Ib}$ & $({\bf 3},{\bf 2},1/6)$ & $({\bf 1},{\bf 2},-1/2)$ & $({\bf 1},{\bf 1},0)$ & -- &  $({\bf 1},{\bf 2 },-1/2)$ \\
\rowcolor{red!35}
$\mathcal{F}_\text{Ic}/\mathcal{S}_\text{Ic}$
& $({\bf 3},{\bf 2},7/6)$ & $({\bf 1},{\bf 2},1/2)$ & $({\bf 1},{\bf 1},-1)$ & $({\bf 1},{\bf 1},0)$ & --  \\
\hline\hline
\rowcolor{red!35}
$\mathcal{F}_\text{IIa}/\mathcal{S}_\text{IIa}$
& $({\bf 3},{\bf 1},2/3)$ & $({\bf 1},{\bf 1},0)$ & $({\bf 1},{\bf 2},-1/2)$ & $({\bf 1},{\bf 2 },-1/2)$ & -- \\
\rowcolor{cyan!40}
$\mathcal{F}_\text{IIb}/\mathcal{S}_\text{IIb}$
& $({\bf 3},{\bf 1},2/3)$ & $({\bf 1},{\bf 1},0)$ & $({\bf 1},{\bf 2},-1/2)$ & --&  $({\bf 1},{\bf 1},-1)$ \\
\rowcolor{red!35}
$\mathcal{F}_\text{IIc}/\mathcal{S}_\text{IIc}$
& $({\bf 3},{\bf 1},-1/3)$ & $({\bf 1},{\bf 1},-1)$ & $({\bf 1},{\bf 2},1/2)$ & --&  $({\bf 1},{\bf 1},0)$ \\
\hline\hline
\rowcolor{red!35}
$\mathcal{F}_\text{Va}/\mathcal{S}_\text{Va}$
& $({\bf 3},{\bf 3},2/3)$ & $({\bf 1},{\bf 1},0)$ & $({\bf 1},{\bf 2},-1/2)$ & $({\bf 1},{\bf 2},-1/2)$ & -- \\
\rowcolor{cyan!40}
$\mathcal{F}_\text{Vb}/\mathcal{S}_\text{Vb}$
& $({\bf 3},{\bf 3},2/3)$ & $({\bf 1},{\bf 1},0)$ & $({\bf 1},{\bf 2},-1/2)$ & --&  $({\bf 1},{\bf 1},-1)$ \\
\rowcolor{red!35}
$\mathcal{F}_\text{Vc}/\mathcal{S}_\text{Vc}$
& $({\bf 3},{\bf 3},-1/3)$ & $({\bf 1},{\bf 1},-1)$ & $({\bf 1},{\bf 2},1/2)$ & --&  $({\bf 1},{\bf 1},0)$ \\
\hline
\end{tabular}
\caption{Minimal sets of fields  fulfilling all requirements listed in the introduction. The fields are denoted by their transformation properties under, respectively, ($SU(3)_c$, $SU(2)_L$, $U(1)_Y$). Models highlighted in cyan feature singlet DM, models in red have singlet-doublet mixed DM.
\label{tab:models}}
\end{table}

\paragraph{Class $\mathcal{S}$.} In these models, we introduce a scalar flavour mediator $\Phi$ and two fermions $\Psi_q$ and $\Psi_\ell$ in vector-like representations of the SM gauge group, plus either 
an additional $\Psi_\ell^\prime$ or a $\Phi^\prime$.
The Lagrangians and the mass matrices are as those given above \emph{mutatis mutandis}:
\begin{eqnarray}
& {\cal L}^{\Psi_\ell \Psi_\ell^\prime}_\mathcal{S} \supset  \Gamma^Q_i\, \bar Q_i \, P_R\Psi_q \, \Phi + \Gamma^L_i\,  \bar L_i\,  P_R \Psi_\ell \, \Phi + \Gamma^E_i\, \bar  E_i\, P_L \Psi^\prime_\ell \, \Phi + \lambda_{H1}  \bar \Psi_\ell\,P_R \Psi^\prime_\ell\, H+ \lambda_{H2}  \bar\Psi_\ell \, P_L \Psi^\prime_\ell \, H + {\rm{h}}{\rm{.c.}}\,, \\
 & {\cal L}^{\Phi \Phi^\prime}_\mathcal{S} \supset \Gamma^Q_i\,  \bar Q_i \, P_R \Psi_q \, \Phi + \Gamma^L_i\, \bar L_i\, P_R \Psi_\ell \, \Phi + \Gamma^E_i\,  \bar E_i\, P_L \Psi_\ell \, \Phi^\prime + a_H  \Phi^\dag \Phi^\prime\,H
+ {\rm{h}}{\rm{.c.}}\,.
\end{eqnarray}
To the best of our knowledge, the model that here we call $\mathcal{S}_\text{Ib}$, belonging to this class, is the only example of this kind of models addressing DM, muon $g-2$ and $B$ anomalies which has been already discussed in the literature in Ref.~\cite{Calibbi:2019bay}.


\section{Results}
\label{sec:results}
In this section we will illustrate the phenomenology of the minimal models introduced above, choosing two examples with distinctive DM candidates: fermionic singlet-doublet DM (model $\mathcal{F}_\text{Ib}$), and real scalar DM (model $\mathcal{F}_\text{IIb}$).
\subsection{$\mathcal{F}_\text{Ib}$: Singlet-doublet fermionic DM}

This model is a generalisation of the model that in Ref.~\cite{Arcadi:2021glq} we called $\mathcal{F}_\text{IA;0}$ with Majorana DM $\Psi=({\bf 1},{\bf 1},0)$ (previously studied  also in Ref.~\cite{Cerdeno:2019vpd}), to which we add a doublet fermion, $\Psi^\prime=({\bf 1},{\bf 2},-1/2)=(\Psi^{\prime\, 0},\Psi^{\prime\,-})$. Thus, the DM candidate arises after EWSB  from the mixing of a singlet Majorana fermion and a Dirac doublet. The Lagragian is the following
 \begin{align}
{\cal L}_{\mathcal{F}_\text{Ib}}~ \supset ~ \Gamma^Q_i\, \bar Q_i\, P_R \Psi \, \Phi_q  + \Gamma^L_i \,  \bar L_i\, P_R  \Psi \, \Phi_\ell  + \Gamma^E_i \, \bar  E_i\, P_L \Psi^{\prime}\cdot \Phi_{\ell}    +  \lambda_{HL} \bar\Psi\,P_L  \Psi^{\prime} \cdot H +  \lambda_{HR} \bar \Psi P_R  \Psi^{\prime} \cdot H  + {\rm{h}}{\rm{.c.}}\,.
\end{align}
where $\Phi_\ell=({\bf 1},{\bf 2},-1/2)$, $\Phi_q=({\bf 3},{\bf 2},1/6)$, and the dot denotes a contraction of the $SU(2)_L$ indexes through $\varepsilon_{ab}=(i\sigma_2)_{ab}$, e.g.~$\Psi^{\prime} \cdot H =\varepsilon_{ab}\Psi^{\prime}_a\, H_b$.

The neutral components of $\Psi^\prime$ and $\Psi$ mix giving three Majorana mass eigenstates by Takagi-diagonalization of a  symmetric mixing matrix with a unitarity matrix $V$
\begin{equation}
{\left( {\begin{array}{*{20}{c}}
\Psi_R^{0\:c}\equiv \Psi_L^0\\
\Psi_{L}^{\prime\, 0}\\
\Psi_{R}^{\prime\,0\:c}
\end{array}} \right)_i} = V_{ij} F_{L,j}^{0}\,,
\quad
V^T \left(
\begin{array}{ccc}
M_{\Psi}   &   \lambda_{HL}v/{\sqrt{2}} &  \lambda_{HR}^*v/{\sqrt{2}} \\
\lambda_{HL}v/{\sqrt{2}} & 0 & M_{\Psi^\prime}\\
\lambda_{HR}^*v/{\sqrt{2}} & M_{\Psi^\prime} & 0\\
\end{array}\right) V = \left(
\begin{array}{ccc}     
m^{F^0}_1 & & \\
& m^{F^0}_2 & \\
& & m^{F^0}_3
\end{array}
\right)\,.
\end{equation}
In an analogous way as what done, e.g., in Ref.~\cite{Calibbi:2015nha}, we will trade in our numerical study the pair $(\lambda_{HL},\lambda_{HR})$ with the pair $(\lambda,\theta)$, defined as:
\begin{equation}
    \lambda_{HL}=\lambda \cos\theta,\,\,\,\,\,\,\lambda_{HR}=\lambda \sin\theta.
\end{equation}
Furthermore we will assume all the parameters to be real.

In terms of the above rotations and mass eigenvalues, the Wilson coefficients for $b \to s \ell \ell$ transitions are:
\begin{eqnarray}
\delta C^9_\mu &=& {\cal N}
\frac{\Gamma^{Q*}_{s}\Gamma^Q_{b}}{32\pi \alpha_{\text{EM}}}
\sum_{i,j=1,2,3}
\frac{V_{1i}V_{1j}}{\left(m_{j}^{F^0}\right)^2} 
\left( 
\vert\Gamma^L_\mu\vert^2 V_{1i}V_{1j} - 
\vert\Gamma^E_\mu\vert^2 V_{2i}V_{2j} \right)
\left( F\left(x_{Q j}, x_{\ell j}, x_{i j}\right) + G\left(x_{Q j}, x_{\ell j}, x_{i j}\right) \right) \,,\nonumber\\
\\
\delta C^{10}_\mu &=& - {\cal N}
\frac{\Gamma^{Q*}_{s}\Gamma^Q_{b}}{32\pi \alpha_{\text{EM}}}
\sum_{i,j=1,2,3}
\frac{V_{1i}V_{1j}}{\left(m_{j}^{F^0}\right)^2} 
\left( 
\vert\Gamma^L_\mu\vert^2 V_{1i}V_{1j} + 
\vert\Gamma^E_\mu\vert^2 V_{2i}V_{2j} \right)
\left( F\left(x_{Q j}, x_{\ell j}, x_{i j}\right) + G\left(x_{Q j}, x_{\ell j}, x_{i j}\right) \right) \,,\nonumber
\end{eqnarray}
where we have defined $x_{Q j} = (m_{\Phi_Q}/m_j^{F^0})^2$, $x_{\ell j} = (m_{\Phi_\ell}/m_j^{F^0})^2$, $x_{i j} = (m_i^{F^0}/m_j^{F^0})^2$, we have introduced the normalization
\begin{equation}\label{eq:Ninv}
{\cal N}^{-1} = \dfrac{4G_F}{\sqrt{2}}V_{tb} V_{ts}^*\,,
\end{equation}
and the loop functions read
\begin{equation}
\begin{aligned}
F(x,y,z) &= 
\frac{x^2 \log (x)}{(x-1) (x-y) (x-z)}+\frac{y^2 \log (y)}{(y-1) (y-x) (y-z)}+\frac{z^2 \log (z)}{(z-1) (z-x) (z-y)} \,,\\
G(x,y,z) &= \frac{2x \log (x)}{(x-1) (x-y) (x-z)}+\frac{2y \log (y)}{(y-1) (y-x) (y-z)}+\frac{2z \log (z)}{(z-1) (z-x) (z-y)} \,.
\end{aligned}
\end{equation}
Notice that, in the absence of the coupling to RH muons or
of the singlet-doublet mixing (i.e.~if $V_{1i}V_{1j}V_{2i}V_{2j}=0$ for any $i,j$), the above contributions give $\delta C^{9}_\mu=-\delta C^{10}_\mu$, as expected. 

The effect of our fields on the muon $g-2$ reads
\beq 
\label{eq:amu_fib}
\Delta a_\mu \simeq 
-\dfrac{m_\mu}{2 \pi^2 \,m_{\Phi_\ell}^2}\,\sum_{j=1,2,3} m_j^{F^0} {\rm Re} \left( \Gamma^L_\mu \Gamma^{E}_\mu V_{1j} V_{2j}\right)  \widetilde H\left(x_{\ell j}^{-1}\right)\,,
\quad\text{with }\,
\widetilde H(x)=\dfrac{x^2 - 1 - 2x \log x}{8 (x-1)^3}\,,
\eeq
where we have shown the dominant chirally-enhanced term only. Subdominant contributions can be found e.g.~in Ref.~\cite{Calibbi:2018rzv,Arnan:2019uhr}.

\begin{figure}[t]
\centering
\includegraphics[width=0.48\linewidth]{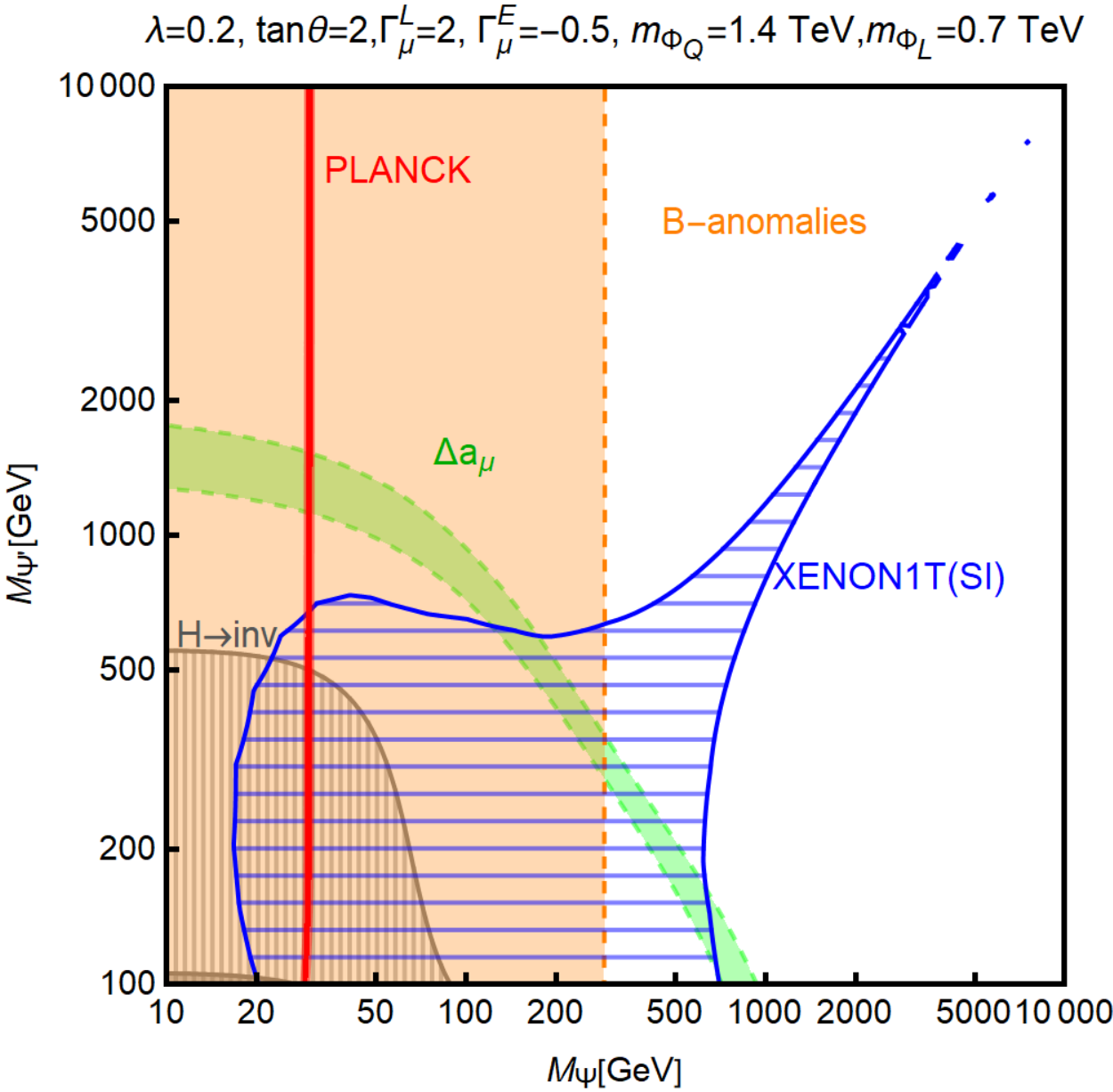}
\hfill
\includegraphics[width=0.48\linewidth]{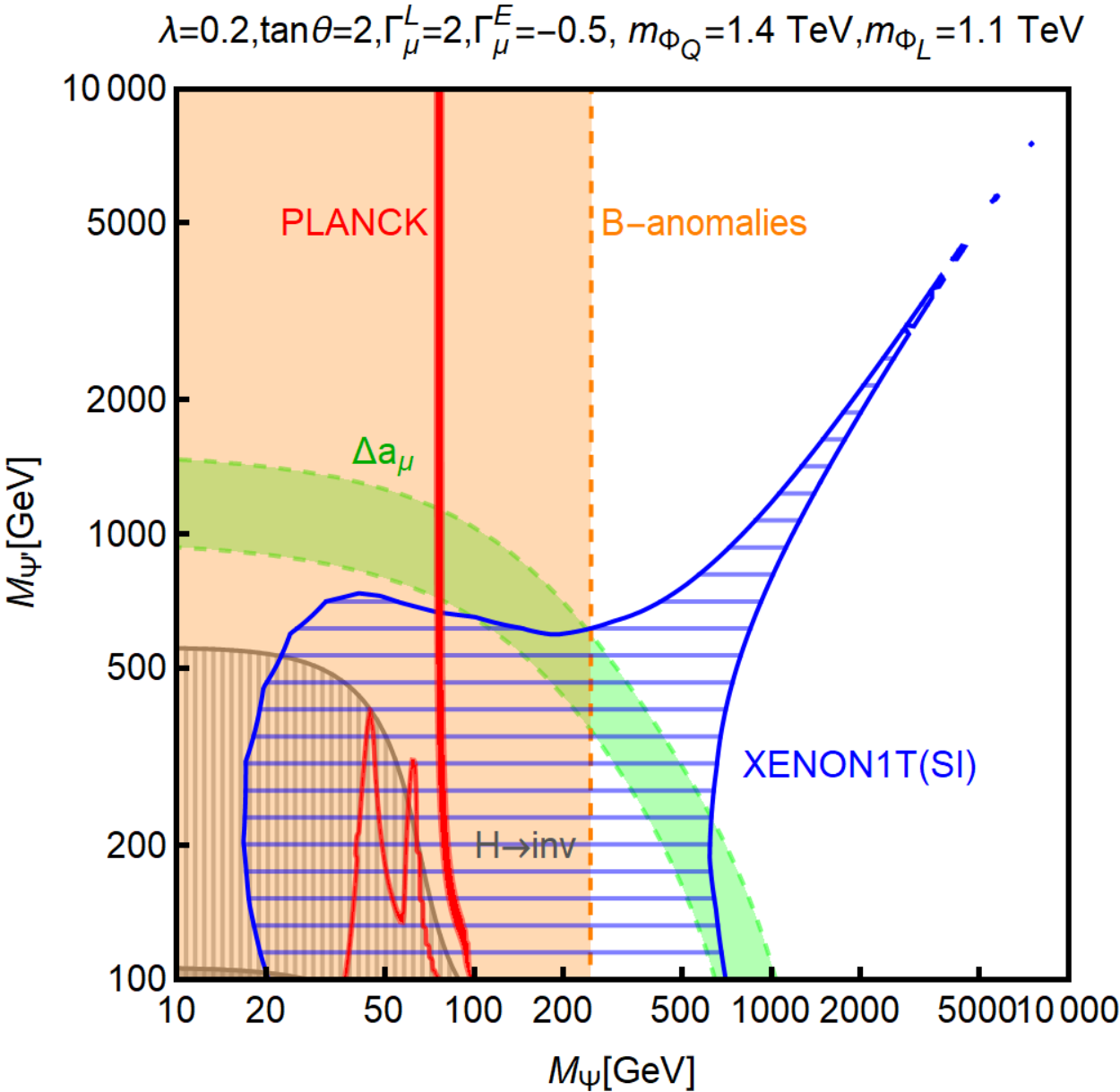}
\caption{Summary of results for the model $\mathcal{F}_\text{Ib}$, in the $(M_{\Psi},M_{\Psi^{'}})$ two-dimensional plane for two assignments of the other model parameters, reported on the top of the two panels. The value of the product of the quark couplings $\Gamma^{Q*}_{s}\Gamma^Q_{b}$ is set to 0.15, according to the constraints imposed by $B_s-\bar{B_s}$ oscillations as found in Ref.~\cite{Arcadi:2021glq}. In each plot the regions of the parameter space accounting for the $g-2$ (at $1\sigma$) and $B$-anomalies (at $2\sigma$) have been marked, respectively, in green and orange. The red isocontours correspond to the correct DM relic density according to the conventional freeze-out paradigm. The blue hatched regions are excluded by Direct Detection limits as given by XENON1T. The gray regions are excluded by searches of invisible decays of the Higgs and $Z$ bosons.}
\label{fig:model1}
\end{figure}

Concerning DM phenomenology, the model discussed here has strong similarities with the model $\mathcal{F}_\text{IA;0}$ illustrated in Ref.~\cite{Arcadi:2021glq}; we will hence refer to the latter work for a detailed discussion. The addition of an $SU(2)_L$ doublet $\Psi^{'}$ causes a notable difference in DM Direct Detection though. In this scenario, indeed, the DM can couple at the tree level with the SM Higgs and $Z$ bosons. These couplings are responsible, respectively, for Spin Independent (SI) and Spin Dependent (SD) interactions between the DM and nucleons, with the former being the most constrained. For a review see e.g. Ref.~\cite{Arcadi:2019lka}. Furthermore, the coupling of DM pairs with the Higgs and the $Z$ bosons would lead to `invisible' decay for the latter bosons, provided that the DM is light enough. Invisible branching fractions for the $Z$ and Higgs bosons are, however strongly constrained.

The results of our analysis are shown in Figure~\ref{fig:model1}, in the $(M_{\Psi},M_{\Psi^{'}})$ two-dimensional plane, for two sample assignations of the parameters of the model. In both panels the green bands represent the regions fitting the muon $g-2$ anomaly at 1$\sigma$ while the regions of the parameter space corresponding to a viable fit of the $B$-anomalies have been marked in orange. Throughout our analyses we set the value of the product of the quark couplings $\Gamma^{Q*}_{s}\Gamma^Q_{b} = 0.15$, in accordance with the constraints imposed by $B_s-\bar{B}_s$ oscillations as found in Ref.~\cite{Arcadi:2021glq}. The correct DM relic density, if conventional freeze-out is assumed, is achieved only in the narrow red strips. The blue-hatched regions are excluded by constraints from XENON1T~\cite{Aprile:2018dbl} on DM SI interactions, while the gray regions corresponds to an invisible branching fraction of the Higgs above $11\,\%$~\cite{Arcadi:2021mag}, or an invisible width of the $Z$ boson above $2.3\,\mbox{MeV}$~\cite{Carena:2003aj}. No analogous constrains from LHC, as the ones considered in~\cite{Arcadi:2021glq}, have been shown in the plot since we have chosen benchmark assignations for $m_{\Phi_{\ell}},m_{\Phi_q}$ beyond current experimental sensitivity. Additional bounds from the production of the charged and neutral partners of the DM should be considered though, being responsible of 2-3 lepton + missing energy signatures (see e.g.~\cite{Aad:2019vvi}). Corresponding limits are not competitive as the ones from Higgs invisible decays and DM direct detection and, hence, have not been shown. 

As illustrated by the figure, a combined fit of the $g-2$ and of the $B$-anomalies can be easily achieved, together with the correct DM relic density and without conflicts with experimental exclusions, for $M_{\Psi}\ll M_{\Psi^{'}}$. This corresponds to a mostly singlet-like DM achieving its relic density mostly through annihilations into muon pairs mediated by $\Phi_\ell$. For this reason the isocontours of the correct relic density appear as vertical lines since the mass of $\Phi_\ell$ and the couplings $\Gamma_\mu^{L,E}$ have been kept fixed in the plots.\footnote{  
Notice that on the right side of the red line (in particular in the region $M_{\Psi}> M_{\Psi^{'}}$ where DM is mostly doublet) our DM candidate is not excluded as it gives a thermal abundance below the observed one. However, some non-thermal DM production mechanism should be at work to account for 100\% of the DM of the universe in this latter case.} It is very promising that both anomalies can be accounted for with a standard thermal DM candidate.

\subsection{$\mathcal{F}_\text{IIb}$: Real scalar DM}
As a second example, we choose a very different DM candidate. Here DM is a real singlet ($\Phi_\ell$): what mix are (the charged components of) the fermion mediator fields $\Psi-\Psi^\prime$ ($SU(2)_L$ doublet and singlet, respectively). Notice that both fields have non-zero hypercharge hence they are both Dirac fermions.
The Lagrangian of the model reads
\begin{align}
{\cal L}_{\mathcal{F}_\text{IIb}} &\supset \Gamma^Q_i\, \bar Q_i \,P_R \Psi \, \Phi_q + \Gamma^L_\mu\, \bar L_\mu\, P_R \Psi \, \Phi_\ell
+ \Gamma^E_\mu\,  \bar E_\mu\, P_L \Psi^\prime \, \Phi_\ell  + \lambda_{HL}  \bar \Psi P_L \Psi^\prime H
+ \lambda_{HR}  \bar \Psi P_R \Psi^\prime H + 
{\rm{h}}{\rm{.c.}}\,,
\end{align}
where $\Phi_q=({\bf 3},{\bf 1},2/3)$, $\Phi_\ell=({\bf 1},{\bf 1},0)$,  $\Psi=({\bf 1},{\bf 2},-1/2)=(\Psi^0,\Psi^-)$  and $\Psi^\prime=({\bf 1},{\bf 1},-1)$. Due to EWSB, $\Psi^\prime$ mixes with the charged state in $\Psi$ and the corresponding mass matrix reads
\beq
\label{eq:MUDE}
\mathcal{M}_\Psi = \left( {\begin{array}{*{20}{c}}
M_{\Psi^\prime}  & \frac{v}{\sqrt{2}} \lambda_{HR}^* \\
\frac{v}{\sqrt{2}}\,  \lambda_{HL} & M_{\Psi} 
\end{array}} \right) \,.
\eeq
We diagonalise this matrix by performing the field rotations
\bea\label{eq:field_rot}
{\left( {\begin{array}{*{20}{c}}
{\Psi^{\prime\,^-}_L}\\
\Psi_{L}^{-}
\end{array}} \right)_i} = U^L_{ij} F_{L,j}^{-}\,,
\qquad
{\left( {\begin{array}{*{20}{c}}
{\Psi^{\prime\,^+}_R}\\
\Psi_{R}^{+}
\end{array}} \right)_i} = U^R_{ij} F_{R,j}^{+}\,,
\qquad
(U^{R\dagger} \mathcal{M}_\Psi U^L)_{ij} = m_{i}^{F^-}\ \delta_{ij}
\eea

The Wilson coefficients of the operators inducing $b \to s \ell \ell$ transitions read:
\begin{eqnarray}
\delta C^9_\mu \!\!&=& \phantom{-}{\cal N}
\frac{\Gamma^{Q*}_{s}\Gamma^Q_{b}}{32\pi \alpha_{\text{EM}}}
\sum_{i,j=1,2}
\frac{U^R_{2i}U^{R*}_{2j}}{\left(m_{j}^{F^-}\right)^2}
\left(
\vert\Gamma^L_\mu\vert^2 U^R_{2i}U^{R*}_{2j}
F\left(x_{Q j}, x_{\ell j}, x_{i j}\right)
- \vert\Gamma^E_\mu\vert^2 U^L_{1i}U^{R*}_{1j}
G\left(x_{Q j}, x_{\ell j}, x_{i j}\right) \right)\,,\nonumber\\
\\
\delta C^{10}_\mu \!\!&=& -{\cal N}
\frac{\Gamma^{Q*}_{s}\Gamma^Q_{b}}{32\pi \alpha_{\text{EM}}}
\sum_{i,j=1,2}
\frac{U^R_{2i}U^{R*}_{2j}}{\left(m_{j}^{F^-}\right)^2}
\left(
\vert\Gamma^L_\mu\vert^2 U^R_{2i}U^{R*}_{2j}
F\left(x_{Q j}, x_{\ell j}, x_{i j}\right)
+ \vert\Gamma^E_\mu\vert^2 U^L_{1i}U^{R*}_{1j}
G\left(x_{Q j}, x_{\ell j}, x_{i j}\right) \right)\,,\nonumber
\end{eqnarray}
where now $x_{Q j} = (m_{\Phi_Q}/m_j^{F^-})^2$, $x_{\ell j} = (m_{\Phi_\ell}/m_j^{F^-})^2$ and $x_{i j} = (m_i^{F^-}/m_j^{F^-})^2$.

The dominant effect on the $(g-2)_\mu$ reads
\beq 
\label{eq:amu_fiib}
\Delta a_\mu \simeq 
\dfrac{m_\mu}{2 \pi^2 \,m_{\Phi_\ell}^2}\,\sum_{j=1,2} m_j^{F^-} {\rm Re} \left( \Gamma^L_\mu \Gamma^{E*}_\mu U^R_{2j} U^{L*}_{1j}\right)  H\left(x^{-1}_{\ell j}\right)\,,
\:\:\text{with}\:\:
H(x)=\dfrac{x^2  - 4x + 3+ 2 \log x}{8 (x-1)^3}\,.
\end{equation}

\begin{figure}[t]
\centering
\includegraphics[width=0.48\linewidth]{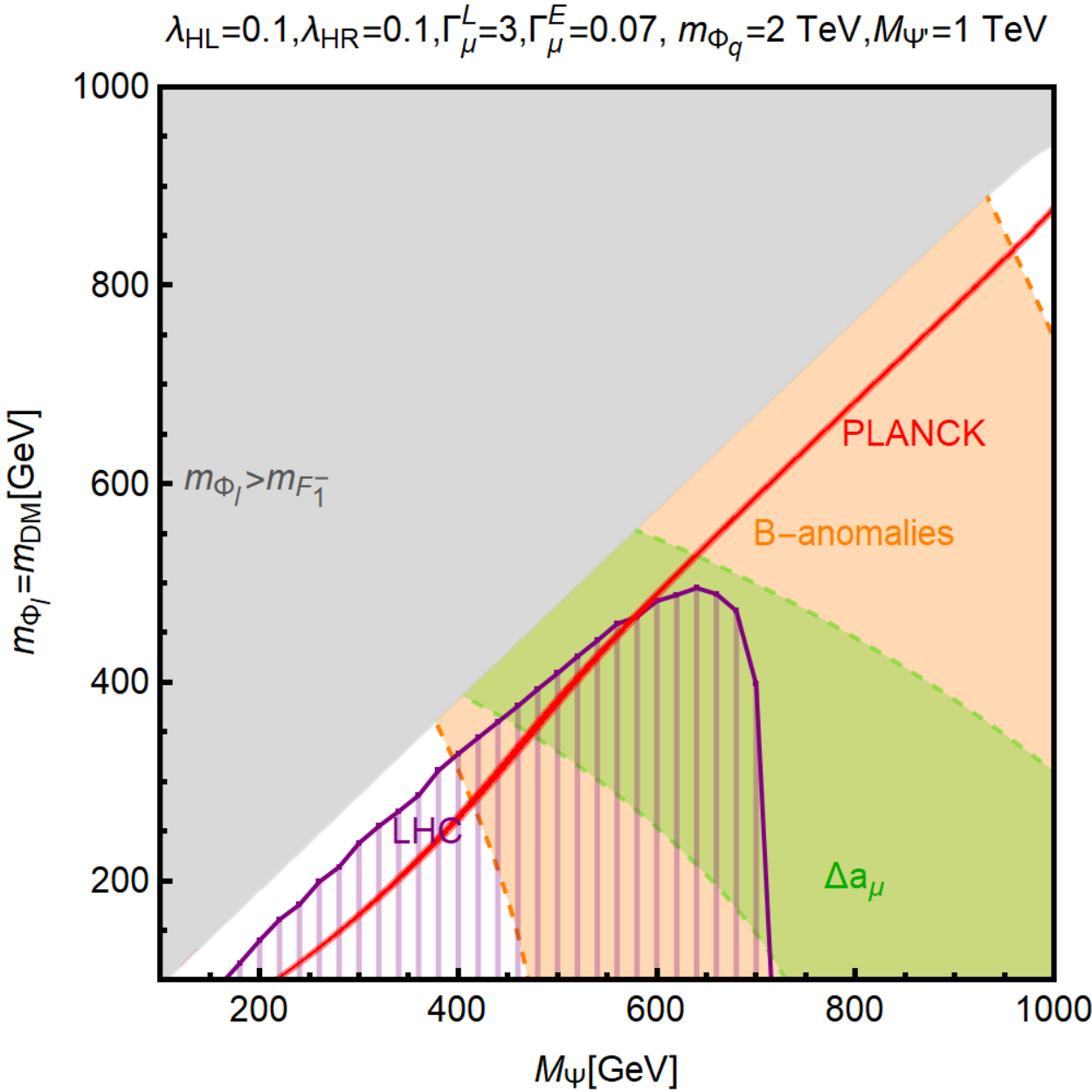}
\hfill
\includegraphics[width=0.48\linewidth]{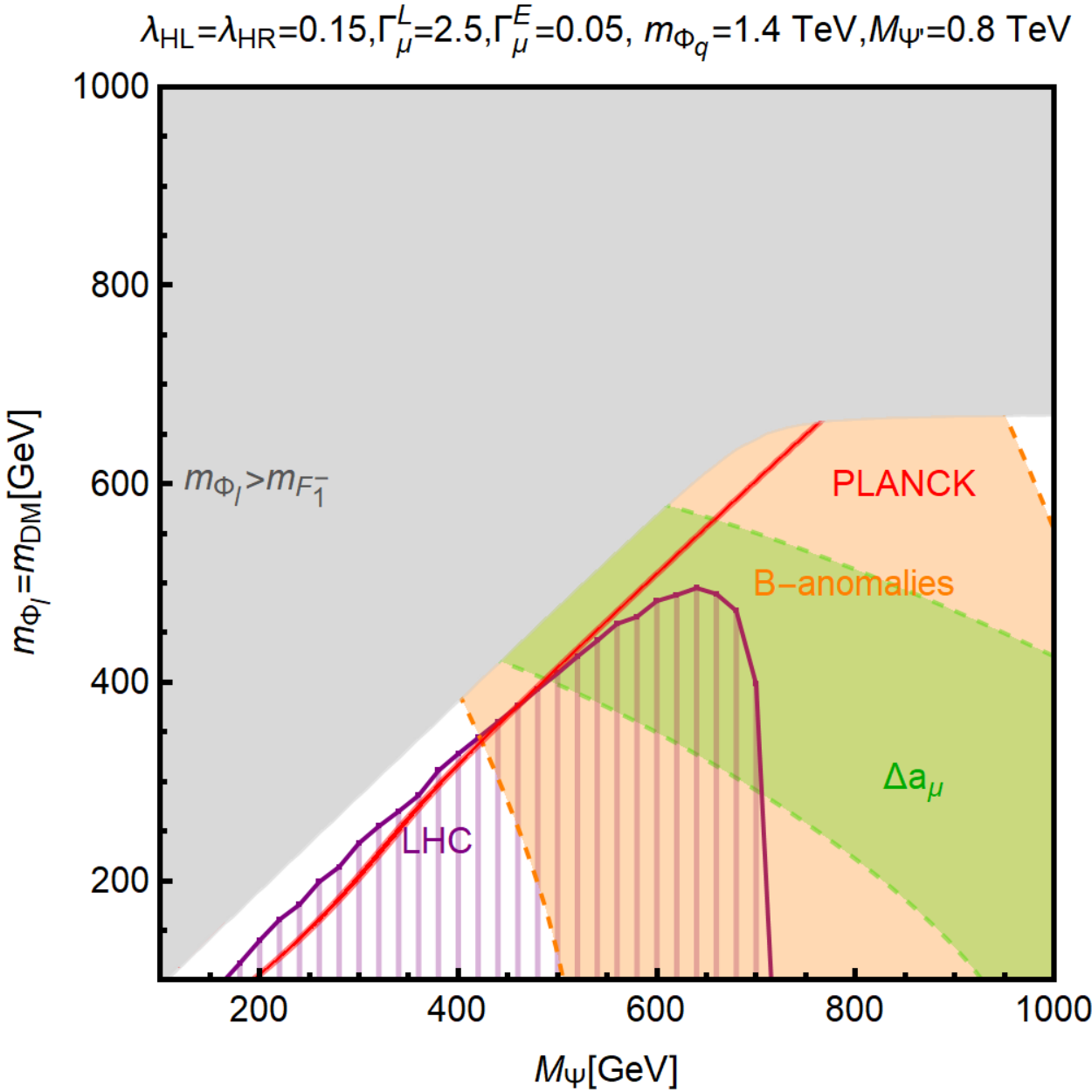}
\caption{Summary of the results for the model $\mathcal{F}_\text{IIb}$ in the $(M_{\Psi},m_{\Phi_\ell})$ two-dimensional plane. The color code is the same as Figure~\ref{fig:model1}.}
\label{fig:model2}
\end{figure}

Again, the details of the DM phenomenology have been fully illustrated in Ref.~\cite{Arcadi:2021glq} and, hence, we will not discuss them in depth here.  
In Figure~\ref{fig:model2}, we show the results of our combined analysis are shown in the $(M_{\Psi},m_{\Phi_{\ell}})$ two-dimensional plane employing the same color coding as in Figure~\ref{fig:model1}. Since, in the figure, the mass of ${F_1^{-}}$ (through $M_\Psi$) is varied, the region of parameter space shown might be impacted by LHC constraints on the production of $F_1^{-}F_1^{+}$ and subsequent decay into $\mu^+\mu^-$ pair and DM, thus missing energy. The corresponding exclusion, obtained by recasting  the search in Ref.~\cite{Aad:2019vvi}, is shown as a  hatched-purple region in the figure. Being the DM a SM singlet real scalar, not interacting with coloured new physics states, the only relevant DM constraint comes from the relic density. Again we found that it is possible to achieve the correct relic density compatibly with a fit of $B-$ and $g-2$ anomalies. 
Again we found that it is possible to achieve the correct relic density compatibly with a fit of $B-$ and $g-2$ anomalies. It is, in particular, worth remarking a special connection between $g-2$ and the DM relic density. Indeed, a real scalar DM interaction, through a fermion mediator, only with left-handed or right-handed fermion would have a d-wave suppressed annihilation cross-section, being the s-wave and p-wave contribution helicity suppressed. In our scenario, the mass mixing between the two fermionic mediators, interacting with left-handed and right-handed muons, removes, at least in part, such helicity suppression~\cite{Berlin:2014tja}. As a consequence, a combined explanation of the anomalies and a thermal DM candidate (not in conflict with any experimental constraint) can be achieved in wide regions of the parameter space.


\section{Conclusions}

The new results presented by the Muon g-2 collaboration could represent the first departure from the prediction of the Standard Model observed in a particle physics experiment. This nicely combines with the growing significance of $R_K$ announced by the LHCb collaboration and the highly-significant deviation from the SM prediction obtained by global fits of all $b\to s\ell\ell$ observables. It is very suggestive that both anomalies requires non-standard contributions to operators involving muons. This makes it crucial to find compelling theoretical frameworks where both experimental results can be naturally explained.
In this paper, we have presented a particularly simple setup where this is possible within the context of models that also provide a viable thermal Dark Matter candidate. We have discussed what are the minimal ingredients and properties that allows to explain the anomalies through loop effects due to the DM particles and few other BSM fields, being four the minimum number of fields that need to be added to the SM.
The general characteristic of this class of models is that the DM phenomenology is controlled by the same parameters that enter the flavour observables. As a consequence, they feature a high degree of correlation among DM, flavour and collider searches and thus an enhanced testability. 
After a systematic classification of these minimal models according to the quantum numbers of their field content, we have presented two examples with very different DM candidates (a mixed $SU(2)_L$ singlet-doublet fermion in one case, a real scalar in the second case). In both examples, a thermal DM candidate can naturally provide a simultaneous explanation of the muon $g-2$ (through chirally-enhanced contributions controlled by EWSB effects) and the $B$-physics anomalies, while evading present bounds from collider and DM searches. Despite the simplicity, their rich phenomenology makes it possible to test them, at least in part, at future runs of the LHC and/or DM direct detection experiments. 
In the long term, as large interactions involving muons are necessary, these minimal DM models would be an ideal target for a multi-TeV muon collider~\cite{Capdevilla:2020qel,Capdevilla:2021rwo,Buttazzo:2020eyl,Yin:2020afe,Chen:2021rnl,Ali:2021xlw}.

\medskip
\noindent {\bf Acknowledgments.}
LC~is partially supported by the National Natural Science Foundation of China under the grant No.~12035008. FM acknowledges financial support from the State Agency for Research of the Spanish Ministry of Science and Innovation through the ``Unit of Excellence Mar\'ia de Maeztu 2020-2023'' award to the Institute of Cosmos Sciences (CEX2019-000918-M) and  from PID2019-105614GB-C21 and  2017-SGR-929 grants. The work of MF is supported by the Deutsche Forschungsgemeinschaft (DFG, German Research Foundation) under grant  396021762 - TRR 257, ``Particle Physics Phenomenology after the Higgs Discovery''.

\bibliography{bibliography}
\bibliographystyle{utphys}

\end{document}